\documentstyle[12pt]{article}
\def\be{\begin{equation}}
\def\ee{\end{equation}}
\def\bea{\begin{eqnarray}}
\def\eea{\end{eqnarray}}

\begin{document}

\begin{center}
{\Large{\bf Cosmological Constant of the $(p+1)$-Dimensional World,
Embedded in the $d$-Dimensional Bulk Space}}

\vskip .5cm
{\large Davoud Kamani}
\vskip .1cm
 {\it Faculty of Physics, Amirkabir University of Technology
\\  P.O.Box: 15875-4413, Tehran, Iran}\\
{\sl e-mail: kamani@cic.aut.ac.ir}
\\
\end{center}

\begin{abstract}

In this manuscript we study the cosmological constant 
of a $(p+1)$-dimensional
world, which lives in the higher dimensional bulk space. We
assume the extra dimensions are compact on tori. We consider
two cases: positive and negative bulk cosmological constant.
It is pointed out that the tiny 
cosmological constant of our world can be 
obtained by the dynamics of a scalar field and 
adjusting the parameters of the model. The cosmological
constant of the dual world also will be discussed. We obtain 
the Dirac quantization of these cosmological constants.

\end{abstract}



\vskip .5cm
\newpage

\section{Introduction}

From the 4-dimensional point of view it seems 
that a solution for the problem of the cosmological constant 
is practically impossible \cite{1}. Therefore,
various attempts have been made to explain the mystery of the cosmological
constant, but no satisfactory solution has been found yet.
In other words, ``in the standard framework of low energy physics there
appears to be no natural explanation for vanishing or extreme smallness of
the vacuum energy'', as Witten put it in \cite{2}.
Braneworlds ($e.g.$ see \cite{3,4}) offer new tools 
that could potentially solve the cosmological constant problem. 
The appearance of extra dimensions in the framework of braneworld
systems seems to provide some new ideas to address this problem from a 
different point of view ($e.g.$ see \cite{3,5} and references therein).
For reviewing the various problems of the cosmological constant 
see \cite{1,6}.

From the other side, a rolling scalar with a nearly flat potential 
provides another useful tool for studying the cosmological constant
\cite{7}. The potential must be very flat 
with some effective continuous variation
in order that the vacuum energy be constant.

In this manuscript we consider a D$p$-brane in the $d$-dimensional bulk space,
with both positive and negative bulk cosmological 
constant, as the general set-up.
Generally, this model is not like the standard braneworld models.
Particular emphasis will be put on the compactification 
of the $d-p-1$ extra dimensions. In our warped model the cosmological constant
of the (p+1)-dimensional universe is not an arbitrary constant.
It is calculable through the moduli of the model.
That is, any physical mechanism which defines (stabilizes) the bulk cosmological
constant, brane tension and radii of compactification together with the
dynamics of the metric of the extra space will 
simultaneously determine the cosmological constant.

We observe that a scalar field, generated by the metric 
of the extra part of the bulk space, generally controls 
the other parameters of the theory. 
That is, when the bulk cosmological constant and the
brane tension go to zero and the 
radius of compactification goes to infinity when the scalar field 
goes to infinity the mechanism of vanishing of the cosmological 
constant generally turns off. 

For matching with the real world, the D3-brane of the type IIB
superstring theory will be emphasized. 
Therefore, the tiny cosmological constant of 
the 4-dimensional world is a consequence of the dynamics of
the scalar field and adjusting of the parameters of the theory.

Finally, cosmological constant of the $(d-p-3)$-dimensional dual
world will be studied. The cosmological constants of the world and its 
dual world obey the Dirac quantization.

This paper is organized as follows. In section 2, the cosmological constant
of the model will be obtained. In section 3,
the cosmological constant of the model with positive bulk cosmological
constant will be analyzed. In section 4,
the cosmological constant due to the negative bulk cosmological constant
will be studied. In section 5, the cosmological constant of the dual world
will be discussed. Section 6 is devoted for the conclusions.
%%%%%%%%%%%%%%%%%%%%%%%%%%%%%%%%%%%%%%%%%%%%%%%%%%%%%%%%%%%%%%%%%%%%%%%%%%%%%%%%%%%%%%%%%%
\section{The cosmological constant of the model}

Consider a D$p$-brane in the $d$-dimensional spacetime (bulk space).
The corresponding action is given by
\bea
S = S_g+S_{DBI},
\eea
where the first term is the $d$-dimensional gravitational action
and the second term is the Dirac-Born-Infeld
(DBI) action corresponding to the D$p$-brane.
Since the Chern-Simons action does not contribute to 
our calculations, we ignore it.

The gravitational action, with the bulk cosmological constant $\Lambda_d$, is
\bea
S_g = \frac{1}{2 \kappa_d^2} \int d^d x \sqrt{-\det G_{\mu\nu}}\;
({\cal{R}}_d- 2\Lambda_d ),
\eea
where $\mu , \nu \in \{0,1,\cdot\cdot\cdot,d-1\}$. 
The $d$-dimensional gravitational constant $\kappa_d$
in terms of the Newton's constant $G_{d}$ is
$\kappa_d^2 = 8\pi G_{d}$. In this action,
up to the cosmological constant $\Lambda_d$, 
we assumed that the graviton is the only bulk field.

For the D$p$-brane the DBI action is
\bea
S_{DBI} = -T_p \int d^{p+1} \xi \sqrt{-\det (g_{\alpha \beta}+
2\pi \alpha' F_{\alpha \beta})}\;,
\eea
where $\alpha , \beta \in \{0,1,\cdot\cdot\cdot,p\}$. Here $g_{\alpha \beta}$ 
is the induced metric on the brane
and $F_{\alpha \beta}$ is the strength of the $U(1)$ gauge field living on the brane.
Since the brane lives in the $d$-dimensional spacetime, 
its tension has the form
\bea
T_p = \frac{\sqrt{\pi}}{2^{(d-10)/4}\kappa_d} (4\pi^2 \alpha')^{(d-2p-4)/4}\;.
\eea

Expansion of the DBI action gives the following terms
\bea
S'_{DBI} = -T_p \int d^{p+1} \xi \sqrt{-\det g_{\alpha \beta}}\;\bigg{(}1+
\frac{1}{4}(2\pi \alpha')^2 F_{\alpha \beta}F^{\alpha \beta}
+\cdot\cdot\cdot\bigg{)} .
\eea 
The second term is the Yang-Mills action in the curved 
background. Therefore, its coefficient defines the 
Yang-Mills coupling. The first term also will be used in calculating
the cosmological constant.
%%%%%%%%%%%%%%%%%%%%%%%%%%%%%%%%%%%%%%%%%%%%%%%%%%%%%%%%%%%%%%%%%%%%%%%%%%%%%%%%%%%%%%%%%%%
\subsection{Compactification of the extra dimensions of the bulk space}

In our model 
the worldvolume of the D$p$-brane describes a $(p+1)$-dimensional
universe. Thus, the bulk space has the $d-p-1$ extra dimensions. 
Assume that all these directions are compact on a tori with equal radii $R$.
In addition, let the fields be independent of the extra coordinates.
Furthermore, we assume that all the brane directions are non-compact.
Therefore, by the standard Kaluza-Klein reduction the action (2) reduces to
\bea
S'_g = &~&\frac{(2\pi R)^{d-p-1}}{2 \kappa_d^2} \int d^{p+1} x 
\sqrt{-\det G_{\alpha\beta}}\;
e^{-2\phi_{p+1}}\bigg{(}{\cal{R}}_{p+1}- 2\Lambda_d 
\nonumber\\
&~& + 4 \partial_\alpha \phi_{p+1}\partial^\alpha \phi_{p+1}
-\frac{1}{4}G^{mn}G^{pq}\partial_\alpha G_{mp}
\partial^\alpha G_{nq} - \frac{1}{4}G_{mn} F^m_{\alpha\beta}
F^{n\alpha\beta}\bigg{)},
\eea
where the indices $m$ and $n$ run over the compact coordinates,
the scalar field is $\phi_{p+1}(x^\alpha) = 
-\frac{1}{4} ln \det G_{mn}$ and $F^m_{\alpha\beta}$ is 
field strength of the gauge bosons $G_{m \alpha}$.
As we see, the scalar field comes from the compactness of the extra dimensions.
In addition, $\phi_{p+1}$ and $G_{mn}$ are not independent fields.
We shall see that this scalar field plays an important role
in the the cosmological constant of the $(p+1)$-dimensional world.

Let the brane be along the non-compact coordinates of the bulk space.
Thus, the embedding functions are $x^\alpha (\xi)=\xi^\alpha$, 
and hence $g_{\alpha \beta}=G_{\alpha\beta}$. 
Besides, we make the following field redefinition
\bea
{\tilde G}_{\alpha\beta} = \exp{\bigg{(}-\frac{4}{p-1}\phi_{p+1}\bigg{)}}
G_{\alpha\beta},
\eea
which is the Weyl transformation. The metric ${\tilde G}_{\alpha\beta}$
is called the Einstein metric. Introducing the metric (7) into
(5) and (6) and summing the resulted actions lead to
\bea
S' = &~& \int d^{p+1} \xi \sqrt{-\det {\tilde G}_{\alpha \beta}}\;\bigg{[}
\frac{(2\pi R)^{d-p-1}}{2 \kappa_d^2}{\tilde {\cal{R}}}_{p+1}  -
\frac{(2\pi R)^{d-p-1}}{\kappa_d^2} \Lambda_d 
\exp \bigg{(}\frac{4}{p-1}\phi_{p+1}\bigg{)}
\nonumber\\
&~& -T_p \exp \bigg{(}\frac{2(p+1)}{p-1}\phi_{p+1}\bigg{)} 
+ \frac{(2\pi R)^{d-p-1}}{2\kappa^2_d} \exp \bigg{(} 
\frac{4}{p-1}\phi_{p+1}\bigg{)}
\nonumber\\
&~& \times \bigg{(} 4 \partial_\alpha \phi_{p+1}\partial^\alpha \phi_{p+1}
-\frac{1}{4}G^{mn}G^{pq}\partial_\alpha G_{mp}
\partial^\alpha G_{nq} - \frac{1}{4}G_{mn} F^m_{\alpha\beta}
F^{n\alpha\beta}\bigg{)}
+\cdot\cdot\cdot \bigg{]},
\eea
where ${\tilde {\cal{R}}}_{p+1}$ is the scalar curvature corresponding 
to the metric ${\tilde G}_{\alpha\beta}$.

The first three terms of 
this action have the feature of the action (2). Thus,
it enables us to read the constants.
The first term defines the gravitational
coupling of the $(p+1)$-dimensional world
\bea
\kappa_{p+1}^2 = (2\pi R)^{p+1-d}\kappa_d^2 . 
\eea
Since $\kappa^2 =8\pi G_{\rm Newton}$, 
the Newton's coupling constants find the relation
\bea
G_{p+1} = (2\pi R)^{p+1-d}G_d .
\eea
In the same way, the second and the third terms give the cosmological 
constant of the $(p+1)$-dimensional world, $i.e.$ $\Lambda^{(d)}_{p+1}$, 
\bea
&~& \Lambda^{(d)}_{p+1} = \Lambda_{\rm bulk}^{(d)} 
+{\bar \Lambda}^{(d)}_{p+1},
\nonumber\\
&~& \Lambda_{\rm bulk}^{(d)}
=\Lambda_d \exp \bigg{(}\frac{4}{p-1}\phi_{p+1}\bigg{)},
\nonumber\\
&~&{\bar \Lambda}^{(d)}_{p+1}= 8\pi G_{p+1}T_p 
\exp \bigg{(}\frac{2(p+1)}{p-1}\phi_{p+1}\bigg{)}.
\eea
The term $\Lambda_{\rm bulk}^{(d)}$ is the cosmological constant
due to the bulk space, and ${\bar \Lambda}^{(d)}_{p+1}$ originates
from the D$p$-brane. Since $\phi_{p+1}$ is a dynamical field, 
the phrase ``cosmological potential'' for $\Lambda^{(d)}_{p+1}$ is more
suitable, but we use the common phrase ``cosmological constant''.

The compactification influences
both components of the cosmological constant. 
The effects of the compact part of the bulk space
are given by $\phi_{p+1}$, and the radius $R$ through the equation (10).

The cosmological constant $\Lambda^{(d)}_{p+1}$ depends on the parameters
$\Lambda_d$, $R$ and $T_p$. Thus, by adjusting these parameters,
the shape of the potential $\Lambda^{(d)}_{p+1}(\phi_{p+1})$ will be fixed.
Then, the dynamical property of $\phi_{p+1}$ enables us to receive the
desired value of $\Lambda^{(d)}_{p+1}$.

The vacuum energy density of the $(p+1)$-dimensional world is as in the following
\bea
&~& \rho_{p+1} =\rho_d(2\pi R)^{d-p-1}
\exp \bigg{(}\frac{4}{p-1}\phi_{p+1}\bigg{)}
+T_p\exp \bigg{(}\frac{2(p+1)}{p-1}\phi_{p+1}\bigg{)},
\nonumber\\ 
&~& \rho_{p+1} \equiv \frac{\Lambda^{(d)}_{p+1}}{8\pi G_{p+1}} 
\;\;,\;\;\rho_d \equiv \frac{\Lambda_d}{8\pi G_d} ,
\eea
where $\rho_d$ is the vacuum energy density of the bulk space. 
We observe that the vacuum energy of the world 
is not sum of the vacuum energies of the brane and bulk.
In addition, finiteness of the vacuum energy density of the world
implies that the extra dimensions have to be compact.
In other words, the radius $R$ cannot go to infinity.
%%%%%%%%%%%%%%%%%%%%%%%%%%%%%%%%%%%%%%%%%%%%%%%%%%%%%%%%%%%%%%%%%%%%%%%%%%%%%
\section{Cosmological constant due to $\Lambda_d > 0$}

We observe that for the positive $\Lambda_d$, 
the world cosmological constant
$\Lambda^{(d)}_{p+1}$ always is a positive quantity.
In other words, positivity of $\Lambda_d$ gives a continues form to
$\Lambda^{(d)}_{p+1}$, from zero to infinity, which are corresponding to
$\phi_{p+1}=-\infty$ and $\phi_{p+1}=+\infty$, respectively.

For sufficiently small $\Lambda_d$ and large extra dimensions, when
$\phi_{p+1}$ has its small values, the cosmological constant is small
and positive. The other interesting case is as follows.
For any values of the parameters, when $\phi_{p+1} \rightarrow -\infty$
the cosmological constant, from the positive side, goes to zero.

Now consider $\Lambda_d \longrightarrow 0$. Therefore, when
the radius of compactification
goes to infinity and the scalar field is at infinity the mechanism of 
vanishing of the cosmological constant generally turns off.

\subsection{Four-dimensional universe}

The D$p$-branes of the superstring theory, $i.e.$ $d=10$, give
\bea
\Lambda^{(10)}_{p+1} = \Lambda_{10}\exp \bigg{(}\frac{4}{p-1}\phi_{p+1}\bigg{)}
+ \frac{g_s}{8 \pi^2 \alpha'}\bigg{(} \frac{\sqrt{\alpha'}}{R} \bigg{)}^{9-p}
\exp \bigg{(}\frac{2(p+1)}{p-1}\phi_{p+1}\bigg{)},
\eea
where $\kappa^2_{10}=8\pi G_{10}=\frac{1}{2} (2 \pi)^7 g_s^2 \alpha'^4$ 
has been used. 

As an interesting example, for the D3-brane of the type IIB
superstring theory the corresponding cosmological constant is
\bea
\Lambda^{(10)}_4 = \Lambda_{10}e^{2\phi_4}
+ \frac{g_s \alpha'^2}{8 \pi^2 R^6}e^{4\phi_4}.
\eea
For an appropriate value of $\phi_4$, $i.e.$ $\phi_4 =\phi_0$,
this should match with the observed cosmological constant
of our 4-dimensional world.
That is, $\Lambda^{(10)}_4 (\phi_0)$ should be positive and small.
This is done by adjusting the moduli $\Lambda_{10}$, $g_s$, $R$
and $\phi_0$. 
In other words, the dynamical property of $\phi_4$ provides a dynamical
mechanism for adjusting the cosmological constant to an almost zero value.
Therefore, in the present time, the universe is in the one of the following
regime: 1) $\phi_4 \rightarrow \phi_0 = -\infty$ and the parameters 
$\Lambda_{10}$, $R$ and $g_s$ are arbitrary and finite;
2) $\phi_4 \rightarrow \phi_0 = 0$ and $\Lambda_{10}$ is small and $R$ is large.
%%%%%%%%%%%%%%%%%%%%%%%%%%%%%%%%%%%%%%%%%%%%%%%%%%%%%%%%%%%%%%%%%%%%%%%%%%%%%%%%%%%%%%
\section{Cosmological constant due to $\Lambda_d <0$}

Let the cosmological constant of the bulk space be negative. 
Therefore, the minimum of the potential (11) takes place at 
$\phi_{p+1}(\xi^\alpha) =\varphi_{p+1}$, where
\bea
e^{2\varphi_{p+1}}=-\frac{\Lambda_d}{4(p+1)\pi G_{p+1}T_p}.
\eea
The minimum of the potential is given by
\bea
\Lambda^{(d)}_{p+1}(\varphi_{p+1})=4\pi (1-p)G_{p+1}T_p
\exp \bigg{(}\frac{2(p+1)}{p-1}\varphi_{p+1} \bigg{)}.
\eea
For $p \geq 2$ this always is nonzero and negative. 
The ratio of the bulk and brane contributions to (16) is
\bea
\bigg{|}\frac{\Lambda_{\rm bulk}^{(d)}}{{\bar \Lambda}^{(d)}_{p+1}}
\bigg{|}_{\phi_{p+1}=\varphi_{p+1}}=\frac{1}{2}(p+1).
\eea
This shows that for $p \geq 2$ the bulk contribution is greater than the
brane contribution.

The potential (11) at ${\bar \phi}_{p+1}$ vanishes, where
\bea
{\bar \phi}_{p+1} = \varphi_{p+1} + \frac{1}{2} ln (\frac{p+1}{2}).
\eea
This leads to the following form for the potential
\bea
\Lambda^{(d)}_{p+1}(\phi_{p+1}) = \frac{8\pi G_d T_p}{(2\pi R)^{d-p-1}}
\exp \bigg{(}\frac{4}{p-1}\phi_{p+1} \bigg{)}
\bigg{(} e^{2 \phi_{p+1}}-e^{2{\bar \phi}_{p+1}} \bigg{)}.
\eea
Therefore, negativity
of $\Lambda_d$ leads to a continues form, from
negative to positive values for the cosmological constant of the
$(p+1)$-dimensional world.
In fact, the dynamics of $\phi_{p+1}$ rolls it toward $\varphi_{p+1}$
and the parameters in the pre-factor of (19) adjust the rapidity of 
this rolling.

According to (4) the potential (19) depends on 
$\sqrt{G_d}$. However,
the large values of $\phi_{p+1}$ generally control small $G_d$ and
large radius of compactification $R$. Thus, the mechanism of vanishing of 
$\Lambda^{(d)}_{p+1}(\phi_{p+1})$ through the variables 
$G_d$ and $R$ generally can turn off by $\phi_{p+1}$.

\subsection{Four-dimensional universe}

Look at the D3-brane of the type IIB superstring theory in the regime
$\Lambda_{10} < 0$. According to the equation (19) 
the cosmological potential of the 4-dimensional world can be written as
\bea
\Lambda^{(10)}_4(\phi_4)=\frac{g_s \alpha'^2}{8 \pi^2 R^6}
e^{2\phi_4}(e^{2\phi_4} - e^{2{\bar \phi}_4}).
\eea
Since the observed value of the cosmological 
constant of the 4-dimensional world is positive,
the scalar field $\phi_4$ of the today's universe is away from
${\bar \phi}_4$. In other words, we have 
$\phi_4|_{\rm present} >{\bar \phi}_4$. 
After this, we receive $\phi_4={\bar \phi}_4$ with $\Lambda^{(10)}_4=0$. 
Then, $\phi_4$ will go toward $\varphi_4$ with negative cosmological constant.
Finally, the universe will obtain the following minimum
\bea
\Lambda^{(10)}_4(\varphi_4)=-\frac{g_s \alpha'^2}{8 \pi^2 R^6}
e^{4\varphi_4}.
\eea
Since $\varphi_4$ is the stable point, the final cosmological
constant of the universe will be negative.

As discussed by Weinberg \cite{8}, a large negative cosmological constant
forces the universe to collapse rapidly. In the same way, 
a large positive cosmological constant causes all matters 
in the universe to disperse. These inspire that the factor
$\frac{g_s \alpha'^2}{8 \pi^2 R^6}$ should sufficiently be small 
to dominate on the $\phi_4$-part of the equation (20).
%%%%%%%%%%%%%%%%%%%%%%%%%%%%%%%%%%%%%%%%%%%%%%%%%%%%%%%%%%%%%%%%%%%%%%%%%%%%%%%%%%%%%%
\section{The cosmological constant of the dual world}

Introducing (4) and (10) into (11) leads to 
\bea
\Lambda^{(d)}_{p+1} = \Lambda_d \exp \bigg{(}\frac{4}{p-1}\phi_{p+1}\bigg{)}
+ \frac{8^{1-d/4}}{\pi^{d/2}}
\sqrt{G_d}\; \alpha'^{(d-2p-4)/4}R^{p+1-d}
\exp \bigg{(}\frac{2(p+1)}{p-1}\phi_{p+1}\bigg{)}.
\eea
This form of the cosmological constant explicitly 
reveals the adjustable parameters 
$\Lambda_d$, $G_d$, $d$, $p$ and $R$ and also dynamical variable $\phi_{p+1}$. 

Now let D$p'$-brane, with $p'=d-p-4$, be dual of the D$p$-brane.
Assume that the D$p'$-brane also is non-compact. 
This implies that the dual brane lives in another
$d$-dimensional bulk space with $d-p'-1=p+3$ compact directions.
According to the number of the compact directions, the bulk spaces
should be different. The cosmological constant of the dual world, for both
cases $\Lambda_d >0$ and $\Lambda_d <0$, is 
given by (22) with $p \longrightarrow p'=d-p-4$,
\bea
\Lambda^{(d)}_{p'+1} = \Lambda_d \exp \bigg{(}\frac{4}{d-p-5}\phi_{p'+1}\bigg{)}
+ \frac{8^{1-d/4}}{\pi^{d/2}}
\sqrt{G_d} \;\frac{1}{\alpha'^{(d-2p-4)/4}}\frac{1}{R^{p+3}}
\exp \bigg{(}\frac{2(d-p-3)}{d-p-5}\phi_{p'+1}\bigg{)}.
\eea

For the self-dual brane, $i.e.$ $p=p'=d/2-2$, both systems are the same.
Therefore, the cosmological constants (22) and (23) reduce to
\bea
\Lambda^{(d)}_{p+1} = \Lambda_d \exp \bigg{(}\frac{8}{d-6}\phi_{p+1}\bigg{)}
+ \frac{8^{1-d/4}}{\pi^{d/2}}
\sqrt{G_d}\; \frac{1}{R^{d/2+1}}
\exp \bigg{(}\frac{2(d-2)}{d-6}\phi_{p+1}\bigg{)}.
\eea 
This depends on the string coupling $g_s$ and string length scale $\alpha'$
through $G_d$. For example, for $d=10$ we have $G_{10}=8 \pi^6 g^2_s \alpha'^4$.
%%%%%%%%%%%%%%%%%%%%%%%%%%%%%%%%%%%%%%%%%%%%%%%%%%%%%%%%%%%%%%%%%%%%%%%%%%%%%%%%%%%%%%%%%%%
\subsection{Dirac quantization of the cosmological constants for
the case $\Lambda_d <0$}

For the R-R charges $\mu_p$ and $\mu_{p'}$ corresponding to the 
D$p$-brane and its dual brane, there is the Dirac quantization
\bea
\mu_p \mu_{p'} = \frac{n\pi}{\kappa^2_d},
\eea
for some integer $n$. The charge $\mu_p$ is related to the D$p$-brane
tension by $\mu_p = g_s T_p$. Thus, according to (19) the product of
$\Lambda^{(d)}_{p+1}$ and $\Lambda^{(d)}_{p'+1}$ 
is a quantized quantity
\bea
\Lambda^{(d)}_{p+1}\Lambda^{(d)}_{p'+1}
= n\frac{8 \pi^2 G_d}{g^2_s(2\pi R)^{d+2}}
\exp \bigg{(}\frac{4}{p-1}\phi_{p+1}+\frac{4}{p'-1}\phi_{p'+1}\bigg{)}
\bigg{(}e^{2\phi_{p+1}} - e^{2{\bar \phi}_{p+1}} \bigg{)}
\bigg{(}e^{2\phi_{p'+1}} - e^{2{\bar \phi}_{p'+1}} \bigg{)}.
\eea 
In the language of energy density, this expression can
be interpreted as the quantization of vacuum energy
density of each world in terms of the unit vacuum 
energy density of its dual world.
%%%%%%%%%%%%%%%%%%%%%%%%%%%%%%%%%%%%%%%%%%%%%%%%%%%%%%%%%%%%%%%%%%%%%%%%%%%%%%%%%%%%%%%%%%
\section{Conclusions and summary}

We derived the effective cosmological constant for 
a $(p+1)$-dimensional world which lives in the $d$-dimensional 
partially compact bulk space.
We observed that finiteness of the vacuum energy density also
imposes the compactness of the extra dimensions.
The effects of the compactification on the bulk and brane
components of the cosmological constant are different. For the
large extra dimensions the brane-component goes to zero.

Since we used the Einstein's frame, a scalar field extracted
from the compactification, appeared in the cosmological constant.
By adjusting the parameters of the model, the
cosmological constant through the dynamics of this scalar field,
goes to the demanded value. For example, the large values of the
above scalar field dominate and hence a nonzero cosmological constant
will be obtained.

For the positive bulk cosmological constant 
$\Lambda_d$, the cosmological constant
of the $(p+1)$-dimensional world is always positive.
In this case, when the scalar field is in $-\infty$, we
obtain $\Lambda^{(d)}_{p+1} \rightarrow 0^+$.
For the negative $\Lambda_d$, the cosmological constant continuously
changes from the positive values to the negative values with a
negative minimum as a stable point. Therefore, in some points it is
positive and very small. In a special point it also vanishes.

We observe that for $\Lambda_d < 0$ the cosmological constants
of the universe and the dual universe obey the Dirac quantization.
In fact, this kind of quantization is incorporated
with the energy quantization.

As an interesting example, for the D3-brane of the type IIB theory, 
the dynamics of $\phi_4$ enables us 
to obtain a tiny cosmological constant of our world for both cases 
$\Lambda_{10}>0$ and $\Lambda_{10}<0$. 
Adjusting the parameters helps us to modify the rapidity of the rolling
scalar $\phi_4$.
However, there exists a solution $\Lambda^{(10)}_4 =0$ 
only for the negative $\Lambda_{10}$.
When the factor $\frac{g_s \alpha'^2}{8 \pi^2 R^6}$ 
is very small, rapid collapse and rapid dispersion do not take place
in the universe.
%%%%%%%%%%%%%%%%%%%%%%%%%%%%%%%%%%%%%%%%%%%%%%%%%%%%%%%%%%%%%%%%%%%%%%%%%%%%%%%%%%%%%%%%%%

\end{document}